\documentclass{emulateapj}
\usepackage{amsmath}
\usepackage{txfonts}
\usepackage[pdftitle={Constraints on ionising photon production from the large-scale Lyman-alpha forest},pdfauthor={Andrew Pontzen}]{hyperref}

\usepackage{xspace}
\usepackage{multirow}
\usepackage{rotating}
\usepackage{amssymb}
\usepackage{comment}
\usepackage[usenames]{color}
\usepackage{filecontents}

\newcommand{\dd}{\mathrm{d}}

\newcommand{\kappaHI}{\kappa_{\mathrm{HI}}}

\newcommand{\deltaGamma}{\delta_{\Gamma}}

\newcommand{\bHI}{b_{\mathrm{HI}}}

\newcommand{\Mpc}{\mathrm{Mpc}}

\newcommand{\HI}{H{\sc i}\xspace}
\newcommand{\hi}{\HI}

\newcommand{\PHI}{P_{\mathrm{HI}}}

\renewcommand{\vec}[1]{\mathbf{#1}}

\newcommand{\deltarho}{\delta_{\rho}}
\newcommand{\bu}{\updated{b_{\mathrm{HI,u}}}}

\newcommand{\xiF}{\xi_{F}}
\newcommand{\xiFl}[1][\ell]{\xi_{F,#1}}
\newcommand{\PF}{P_{F}}
\newcommand{\PFl}{P_{F,\ell}}
\newcommand{\matr}[1]{\mathsf{#1}}
\newcommand{\bFu}{b_{F}}
\newcommand{\bFv}{b_{Fv}}
\newcommand{\bFg}{b_{F\Gamma}}

\newcommand{\updated}[1]{#1}

\begin{document}

\submitted{Published in ApJL, August 27, 2014}

\title{Constraints on ionising photon production from the large-scale Lyman-alpha forest}

\author{Andrew Pontzen$^1$, Simeon Bird$^2$, Hiranya Peiris$^1$, Licia Verde$^{3,4}$}
\affil{$^1$Department of Physics and Astronomy, University College
  London, Gower Street, London WC1E 6BT, UK}
\affil{$^2$Institute for Advanced Study, 1 Einstein Drive,
  Princeton, NJ, 08540, USA}
\affil{$^3$ICREA \& ICC, Institut de Ciencies del Cosmos,
  Universitat de Barcelona (IEEC-UB), Marti i Franques 1, Barcelona
  08028, Spain}
\affil{$^4$Institute of Theoretical Astrophysics, University of Oslo, Norway}

\begin{abstract}
  Recent work has shown that the $z \simeq 2.5$ Lyman-alpha forest on
  large scales encodes information about the galaxy and quasar
  populations that keep the intergalactic medium photoionized.  We
  present the first forecasts for constraining the populations with
  data from current and next-generation surveys. At a minimum the
  forest should tell us whether galaxies or, conversely, quasars
  dominate the photon production. The number density and clustering
  strength of the ionising sources might be estimated to sub-$10\%$
  precision with a {\tt DESI}-like survey if degeneracies ({\it e.g.},
  with the photon mean-free-path, small-scale clustering power
  normalization and potentially other astrophysical effects) can be
  broken by prior information. We demonstrate that, when inhomogeneous
  ionisation is correctly handled, constraints on dark energy do not
  degrade.
\end{abstract}

\maketitle

\section{Introduction}

Almost fifty years after \cite{1965ApJ...142.1633G} first used quasar
spectra to infer the intergalactic neutral hydrogen density, the
Lyman-$\alpha$ forest remains a key probe of cosmological physics. The
forest has been widely used to measure small-scale structure
\cite[\textit{e.g.},][]{Croft99,McDonald_SDSS_1DLyA_Pk_2006} and more
recently datasets of tens of thousands of spectra have allowed us to
correlate cosmic density over tens to hundreds of megaparsecs
\citep{McDonald03,McQuinnWhite11_LyA_estimators,BOSS_Slosar11}. This
opens up the possibility of constraining the expansion history --
and so dark energy -- by measuring the scale of the baryon acoustic
oscillations (BAO)
\citep{McDonaldEisenstein07_LyABAOforecast,2013_Busca_BOSS_Lya_BAO,BOSS_Slosar13,Delubac14_Lya_BOSSDR11}.
But there is more information to be harvested given a sufficiently
accurate measurement of the Lyman-$\alpha$ correlation function. The
possibility that astrophysics distorts cosmological correlation
functions has been foreseen for some time \citep[{\it
  e.g.},][]{Bower93_cooperativeGform} but concrete processes have
typically been studied only over quite small patches of a few
megaparsecs \cite[{\it
  e.g.},][]{Kollmeier03_LyA_LBG_connection,Meiskin04_LyA_radiation}. We
focus here on a correction to fluctuations in patches of size tens to
hundreds of megaparsecs, arising from a radiation effect that
increases in amplitude as the scales become comparable to the mean
free path of an ionising photon
\citep{Croft04_LyA_large_radiation,McDonald05,McQuinn_11_LyLimit,Pontzen14_BAObias,Gontcho14}. We
will use the description of \citeauthor{Pontzen14_BAObias} (2014,
henceforth P14) to demonstrate that one can constrain the nature of
ionising radiation sources in the redshift range $2<z<3$, provided
that various calibration uncertainties are under control.

The underlying physical process is an analogue of the proximity effect
\citep[\textit{e.g.},][]{1986ApJ...309...19M,Batjlik88_proximity,Dallaglio_08_proximity,Calverley_11_proximity}
-- in regions of high ultraviolet (UV) photon density, the forest is
suppressed because the neutral fraction of hydrogen declines. But
while the classical proximity effect applies in the relatively small,
several-megaparsec region where a single nearby quasar appreciably
boosts the UV background, the process here arises due to averaged
fluctuations in emissivity over much larger scales. The
finite mean free path dictates that ionising radiation cannot reach
uniformity in patches larger than a few hundred comoving megaparsecs;
if one probes \hi fluctuations on scales approaching this limit, there
are unavoidable distortions in the power spectrum.

The P14 analysis solves the Boltzmann radiative transfer equation by
assuming that the radiation field is in local equilibrium, and that
the equations can be linearized by averaging over small
scales. The result is a quantitative link between the power spectrum of
\hi fluctuations ($\PHI(k)$) and that of the total density ($P(k)$),
dependent on a number of astrophysical parameters.  These include the
bias of UV sources, $b_j$; the number density of UV sources, $\bar{n}$;
and the opacity of the intergalactic medium to Lyman-limit ({\it
  i.e.}, ionising) radiation, $\kappaHI$. When multiple populations
contribute, $b_j$ and $\bar{n}$ average in well-defined ways. As a reference model, we will adopt the P14 default parameter
values: $b_j \simeq 3$, $\bar{n} \simeq 10^{-4}\,h^{3}\,\Mpc^{-3}$ and
$\kappaHI\simeq (390\,h^{-1}\,\Mpc)^{-1}$.

The detailed calculation given by P14 reveals that, even though source
clustering is weak over the vast distances involved, radiative
transfer generates a major correction to the expected power spectrum
because the clustering of the gas is weaker still. Shot noise is also
a major factor: the rarity of sources, again a small effect on such
scales, can still be significant compared to the tiny cosmological
clustering power. These results are reinforced by an alternative
calculation by \cite{Gontcho14}.  There are, however, potential
complications arising from temperature fluctuations. These impact on
the flux transmission by modifying the neutral fraction and by
changing the shape of a cloud's absorption profile; they were not
included in the analysis of P14, but an estimate by \cite{Gontcho14}
showed them to generate only a small correction to the power
spectrum. We therefore continue to ignore them for our exploratory
work here (but see Section~\ref{sec:conclusions}).

\section{Correlation functions and their parameters}\label{sec:params}

To make forecasts for future forest analyses, we constructed a
pipeline based on the BOSS (Baryon Oscillation Spectroscopic Survey)
approach as described by \citeauthor{2013_Busca_BOSS_Lya_BAO} (2013,
henceforth B13). While P14 gives a prediction for the linear \hi power
spectrum, $\PHI(k)$, here we need the Lyman-alpha transmission flux
power spectrum $\PF(k)$. We consider only scales that are far above
the non-linear dynamics regime; on the other hand the non-linear
effects on small scales still cannot be entirely neglected \citep[{\it
  e.g.},][]{McDonald00,McDonald03,Seljak12LyaBias}. In particular
spatial averaging does not commute with the transformation to flux,
and our $\PHI(k)$ therefore cannot simply be rescaled. Instead one
must decompose the processes shaping $\PHI(k)$ back into their
separate physical origins and reassemble them in an appropriate way as
we now describe. This process will also introduce angle-dependence
from redshift-space distortions
\citep{1987MNRAS.227....1K,Hamilton98_RSDreview}.

In the absence of radiation fluctuations one has two parameters: a
density bias $\bFu$ and a velocity bias $\bFv$.  These relate changes
in the large-scale average flux field to the linear fractional
overdensity field $\deltarho$ by
\begin{equation}
\delta_F = \left[\bFu +  (1+\mu^2)\bFv\right] \deltarho\textrm{,}
\end{equation}
where $\delta_A=A(\vec{x})/\langle A \rangle -1$ for any field $A$,
and $\mu$ is the cosine of the angle to the line-of-sight. The
dependence of the redshift-space distortions on cosmology has been
neglected since we are working at high redshift
\citep{Hamilton98_RSDreview}. From simulations
\cite[\textit{e.g.},][]{McDonald03} one has $\bFu \simeq -0.14$ and
$\bFv \simeq -0.20$, compatible with the observational constraint on
$\bFu+\bFv \simeq -0.34$ \citep{BOSS_Slosar11}.

When we introduce inhomogeneous radiation, a further term must be added:
\begin{equation}
\delta_F = \left[ \bFu  +  (1+\mu^2) \bFv \right]\deltarho + \bFg \deltaGamma\textrm{,}\label{eq:deltaF}
\end{equation}
where $\deltaGamma$ are the fractional fluctuations in the ionisation
rate and $\bFg$ is a new bias parameter describing the response of the
flux. The redshift space distortions, being gravitational in origin,
are unaffected by inhomogeneous radiation at first order ($b_{Fv}$ is
unchanged). A numerical value of $\bFg$ is determined by averaging
over the response of individual Lyman-$\alpha$ lines to an increase in
\hi fraction \citep{FontRibera_BOSS_QSO_Lya_X_2013}, giving $\bFg
\simeq 0.13$ \citep{FontRibera_BOSS_QSO_Lya_X_2013,Gontcho14}. We
also verified this result using the simulations of \cite{Bird11_Lya}.

The radiation fluctuations correlate with the
cosmic density so that in Fourier space
\begin{equation}
\langle \deltaGamma(k) \deltarho(k) \rangle \propto b_{\Gamma\rho}(k) P(k)\textrm{,}
\end{equation}
where $b_{\Gamma\rho}(k)$ is the scale-dependent bias describing the
relationship between radiation and $P(k)$, the underlying cosmological
power spectrum. The bias $b_{\Gamma\rho}(k)$ encodes the radiative
transfer physics and scales near-linearly with the source clustering
strength $b_j$ [in P14 $b_{\Gamma\rho}(k)$ is given by equation (35)
since $\bHI(k) = \bu - b_{\Gamma\rho}(k)$ in the notation there].  An
additional, uncorrelated shot noise contribution enters so that
\begin{equation}
\langle \deltaGamma(k) \deltaGamma(k) \rangle \propto b_{\Gamma\rho}(k)^2 P(k) + N(k)\textrm{.}
\end{equation}
The scale-dependence of $N(k)$ is specified by the radiation transfer
physics [corresponding to the second term in equation (38) of P14];
its amplitude is inversely proportional to the number density
$\bar{n}$ of sources.

\begin{figure}[t]
\includegraphics[width=0.5\textwidth]{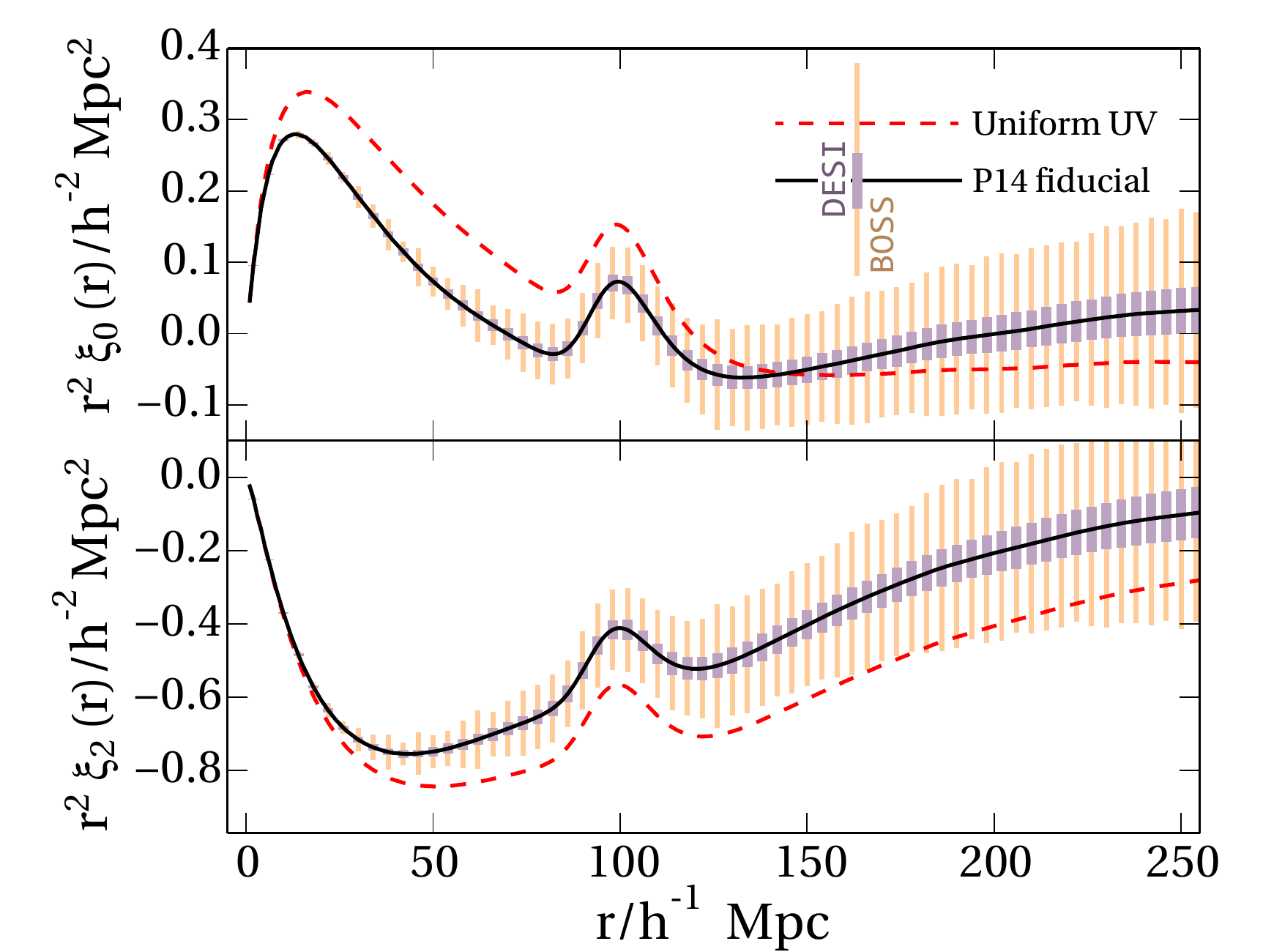}
\caption{The first two multipole moments of the redshift-space flux
  correlation function $\xi_0$ and $\xi_2$; the dashed line
  corresponds to a uniform UV background, and the solid line to the
  P14 solution with UV from a mix of quasars and galaxies. As the
  dominant sources change from galaxies to quasars, the solution moves
  further away from the uniform limit (see figures 3 to 5 in P14). The
  thick and thin error bars show respectively the diagonal part of the
  covariance for \texttt{DESI} and \texttt{BOSS}. Correlation of the
  errors is $\simeq 20\%$ between neighbouring
  measurements.}\label{fig:multipoles}
\end{figure}

Combining the equations above, the flux power spectrum is
\begin{equation}
P_F(k,\mu) = \left[\bFu + (1+\mu^2)b_{Fv} + \bFg b_{\Gamma\rho}(k) \right]^2 P(k) + \bFg^2 N(k)\textrm{.}\label{eq:main-powspec}
\end{equation}
Current analyses of the forest measure
the correlation function $\xiF(r,\mu) = \langle \delta_F(\vec{x}) \delta_F(\vec{x}+
\vec{r}) \rangle$, where $\mu$ is the angle between the displacement
vector $\vec{r}$ and the line of sight. This contains equivalent
information to the power spectrum $P_F(k,\mu)$ and is related by
decomposing the angular dependence into Legendre polynomials:
\begin{equation}
P_F(k,\mu) = \hspace{-0.2cm} \sum_{\ell=0,2,4} \PFl(k) p_{\ell}(\mu)\textrm{;}
\hspace{0.2cm} \xiF(r,\mu) = \hspace{-0.2cm} \sum_{\ell=0,2,4} \xiFl(r)
p_{\ell}(\mu)\textrm{,} \hspace{-0.05cm}
\end{equation}
where $p_{\ell}$ is the Legendre polynomial of order $\ell$. (The
linear-order \citeauthor{1987MNRAS.227....1K} 1987 approximation only generates
terms with $\ell=0,2$ and $4$.)  Calculating the moments of the power
spectrum $\PFl(k)$ is a matter of rewriting the $\mu$-dependence of
\eqref{eq:main-powspec} in terms of the $p_{\ell}$'s. The final step
is to relate $\xiFl$ to $\PFl$; one may show that
\begin{equation}
\xiFl(r) = \frac{i^{\ell}}{2 \pi^2} \int_{0}^{\infty} \dd k\, k^2 \PFl(k) j_{\ell}(kr)\textrm{,}
\end{equation}
where $j_{\ell}$ is the spherical Bessel function of order $\ell$
\citep{Hamilton98_RSDreview}. Following B13 we will work directly with
the multipoles. An example of $\xiFl[0]$ and $\xiFl[2]$ is shown in
Figure~\ref{fig:multipoles} for our standard parameters (solid line)
and the equivalent model with a completely uniform UV background
(dashed line). Cosmological parameters are unchanged from P14 and
based on \cite{Planck13_CosmoPar}; the underlying power spectrum is
calculated with CAMB \citep{Lewis:1999bs}.

The Lyman-$\alpha$ forest constrains the angular diameter distance and
Hubble parameters via the transverse and line-of-sight BAO.  Current
analyses (\textit{e.g.}, B13) assume a fiducial cosmology during an
initial conversion from raw data to the correlation function, then
measure departures of the BAO feature from its expected location.  We
therefore decompose all our power spectra into two parts, a `smooth'
and a BAO `peak' component, following the recipe given by
\cite{Kirkby13Lya}. The peak component is then shifted in scale by a
fixed factor $\alpha$; for $\alpha=1$, one recovers the exact fiducial
cosmology.
If one fixes the transfer function, the forest encodes further
information on the spectral index of primordial fluctuations $n_s$; we
add another proxy cosmological parameter, $\zeta$, such that
\begin{equation}
  P_F(k) \to P_F(k) \left(\frac{k}{0.1\,h^{-1}\,\Mpc} \right)^{\zeta-1}\textrm{.}\label{eq:define-zeta}
\end{equation}
This definition returns the default cosmology when $\zeta=1$ and
pivots around the arbitrary scale $0.1\,h^{-1}\,\Mpc$. In a realistic
case $\zeta$ will constrain a degenerate combination of $n_s$ and
cosmic density parameters.

\begin{table}
\begin{tabular}{ll}
\hline
& {\it Cosmological parameters} \\
$\alpha$ & Scaling of the BAO peak position \\
$\zeta$ & Broadband spectral tilt $=n_s$ if transfer
function is fixed \\
\hline
& {\it Astrophysical parameters (scale-independent)} \\
$b_j$ & Bias of UV sources (weighted average) \\
$\bar{n}$ & Number density of UV sources (weighted average) \\
$\kappaHI$ & Opacity of the intergalactic medium to ionising
radiation\\
$\bu$ & Bias of intergalactic \hi ignoring UV fluctuations \\
\hline
& {\it Astrophysical functions (scale-dependent)} \\
$b_{\Gamma\rho}(k)$ & Bias of UV fluctuations, depends on $b_j$ and $\kappaHI$ \\
$N(k)$ & Noise from UV fluctuations, depends on $\bar{n}$ and $\kappaHI$ \\
$\bHI(k)$ & = $\bu - b_{\Gamma\rho}(k)$; overall bias of intergalactic \hi \\
\hline
& {\it Forest parameters (scale-independent)} \\
$\bFu$ & Bias of the observed forest (ignoring UV and velocity) \\
$\bFv$ & Bias of the observed forest relative to redshift-space distortions \\
$\bFg$ & Bias of the observed forest relative to UV intensity  \\
\hline
& {\it Nuisance parameters} \\
$a_{\ell,n}$ & Six flux-calibration broadband distortion parameters \\

\hline
\end{tabular}
\caption{A reference guide for the quantities used in this Letter.}\label{tab:parameters}
\end{table}

Finally, following B13, we introduce nuisance parameters to
characterise distortions induced by current pipelines:
\begin{equation}
\xiFl \to \xiFl + \sum_{n=0}^2 a_{\ell,n} r^{-n}\textrm{.}
\end{equation}
These have been shown by use of mocks to be an adequate description of
uncertainty from quasar continuum estimates relying on data in the
region of the forest \cite[][B13]{Kirkby13Lya}.



\section{Statistical techniques and cosmological biases}

We constructed covariance matrices $\matr{C}$ summarising the expected
noise properties of two separate survey configurations. We considered
only those data lying in the redshift range $2<z<3$; the two major
parameters are then the number of quasars $N_{\mathrm{QSO},z}$ in this
redshift range and the sky coverage area $A$. We refer to our
covariances as {\tt BOSS}, corresponding to the final planned data
release ($N_{\mathrm{QSO},z}=1.5\times 10^5$, $A=10\,000$ sq. deg.:
\citeauthor{2013Dawson_BOSS_overview} 2013); and {\tt DESI}, referring
to a futuristic survey modelled on the Dark Energy Spectroscopic
Instrument ($N_{\mathrm{QSO},z}=7.5\times 10^5$, $A=14\,000$ sq. deg.:
\citeauthor{Levi13_DESI} 2013). We made use of the recipe given by B13
to generate the diagonal and leading-order off-diagonal terms from an
estimation of $N_{\mathrm{pair}}$, the number of survey pixels
separated by a given spatial distance. This will be a reasonable
approximation independent of cosmology (since the dominant terms arise
from one-point variance related to local, rather than large-scale,
structure) and independent of survey (because the target
signal-to-noise is similar for {\tt BOSS} and {\tt DESI}).
These considerations also explain why the
covariance matrix is near-diagonal.
To verify our $N_{\mathrm{pair}}$ estimation
procedure we emulated {\tt BOSS-DR9}, based on numbers quoted
in B13 ($N_{\mathrm{QSO},z}=4.8\times 10^4$, $A=3\,300$ square
degrees) and verified that our estimated covariance matrix very
closely mimics the published bootstrap estimation by B13.

Results will be quoted using only separation scales larger than
$40\,h^{-1}\,\Mpc$, to ensure that the linear approximations of P14
hold ({\it i.e.}, that local non-linear physics is safely segregated in
constants such as  $\bu$, $b_v$ and $\bFg$). We assumed uniform priors
on all parameters and adopted a Gaussian likelihood approximation,
equivalent to B13's use of $\chi^2$ statistics.

\begin{figure*}
\begin{center}
  \includegraphics[width=0.9\textwidth]{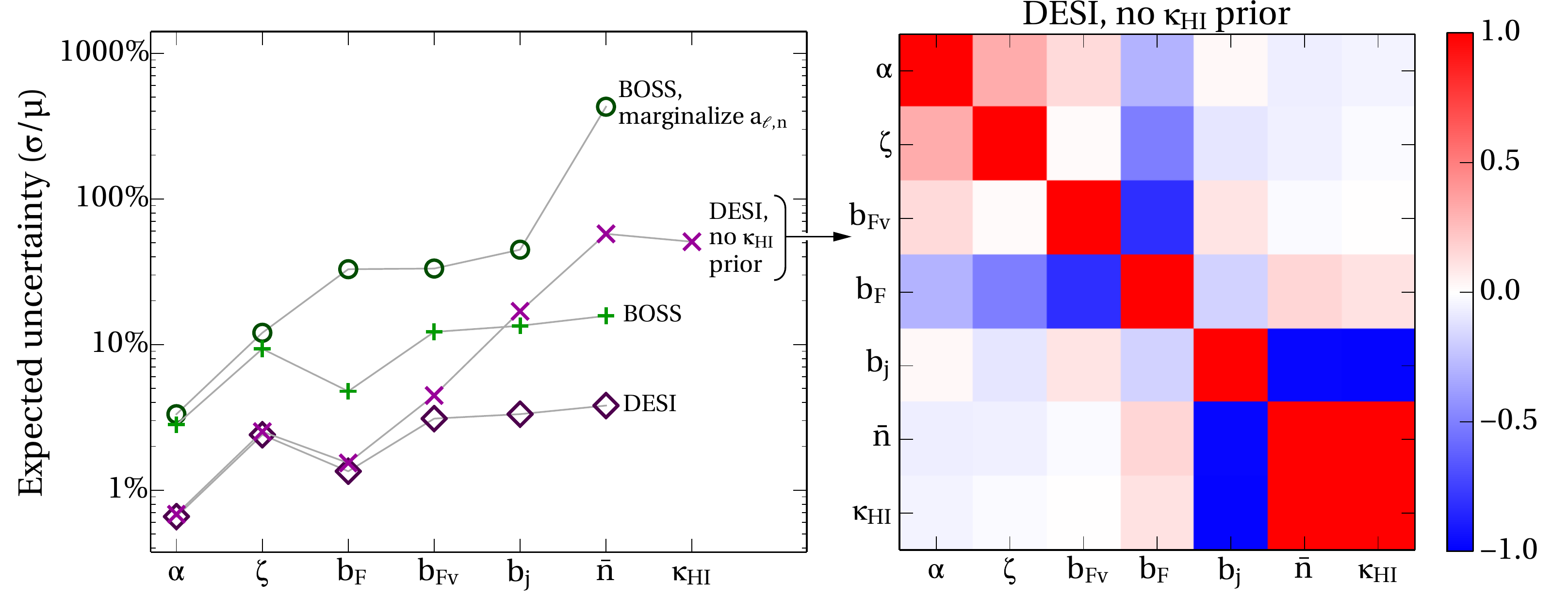}
\end{center}
\caption{({\it Left panel}) the forecast relative errors ({\it i.e.},
  the rms deviation divided by the mean) for simultaneously
  constraining all parameters with {\tt BOSS} or {\tt DESI}-like
  surveys.  See Table \protect\ref{tab:parameters} for an explanation
  of each variable. With {\tt BOSS} ($+$ symbols) and {\tt DESI}
  ($\diamond$ symbols) one can derive significant constraints on $b_j$
  and $\bar{n}$ which describe the origin of UV photons, although
  these are weaker if the intergalactic medium opacity $\kappaHI$ is
  assumed unknown ($\times$ symbols). Traction on the astrophysics is
  lost if the continuum-induced broadband distortions are not
  adequately characterised ($\circ$ symbols). ({\it Right panel}) the
  correlation matrix for {\tt DESI}-like case shows that the error
  budget for $b_j$ and $\bar{n}$ arises largely from a degeneracy with
  $\kappaHI$, explaining the major difference between $\times$ and
  $\diamond$ cases in the left panel.}\label{fig:fisher}
\end{figure*}

We first investigated whether cosmological parameter estimation may be
biased by the inhomogeneous radiation. This is a matter
of generating mock data using our fiducial UV solution, but attempting
to fit the results without any radiative distortions ({\it i.e.}, with
$\bFg=0$), leaving the nuisance parameters $a_{\ell,n}$ to mop up the
resulting broadband distortion. Current pipelines closely recover the
BAO peak position despite the unmodeled effects. In particular using
the {\tt BOSS} covariance matrix, we found $\alpha=1.004\pm 0.025$
where the central value is the marginalized mean over the posterior
and the error is its standard deviation. For {\tt DESI}, we obtained
$\alpha = 1.003 \pm 0.005$.

These expected systematic errors are significantly smaller than the
$1\%$ shift in the BAO peak reported in P14 for two reasons: first,
much of the observed signal comes from the redshift space distortions
(which are not subject to radiative effects). Second, P14 directly
measured the position of the peak, whereas here we marginalized over a
generic set of broadband distortions $a_{\ell,n}$ which partially remove the shape
modulation.

Consequently the expected biases are an order-of-magnitude too small
to account for a several-percent level tension with $\Lambda$CDM
reported by \cite{Delubac14_Lya_BOSSDR11}.  Even if one substantially
increases the size of the modulation, for instance by reducing
$\bar{n}$ or increasing $b_j$, the $a_{\ell,n}$-fitting restricts the
errors to below one percent. On the other hand with future surveys
this decoupling may not be sufficient -- for example with {\tt DESI}
and $b_j=4$ (reflecting a large quasar contribution) we find $\alpha =
1.006 \pm 0.006$, which starts to be problematic.

Errors can be reduced by implementing a more flexible broadband
distortion function; or they can be eliminated by fitting
astrophysical parameters. The latter approach delivers insight into
the galaxy and quasar population, as we now discuss.

\section{Astrophysical forecasts}\label{sec:astr-forec}

We use a Fisher matrix formalism \cite[see, {\it e.g.},][]{DETF06} to
estimate the covariance matrix for maximum likelihood parameter
estimates, determining how accurately astrophysical and cosmological
parameters can be inferred.  Figure~\ref{fig:fisher} shows the results
for simultaneously fitting seven key parameters from Section
\ref{sec:params} (see Table \ref{tab:parameters}).  In the left panel
we show the diagonal elements of the Fisher matrix divided by the true
value of the parameter: this gives a prediction for the $1\sigma$
relative error for a given quantity, marginalized over the other six
parameters. The different point styles represent different scenarios
described below.

Consider the most optimistic scenario -- a {\tt DESI}-like survey
with accurate broadband characterisation, plus a known intergalactic
medium optical depth $\kappaHI$ to ionising photons ($\diamond$
symbols, left panel of Figure \ref{fig:fisher}). In this case, one
obtains a constraint on~$b_j$, the clustering bias of UV sources, and
on $\bar{n}$, the number density of such sources, both accurate to
around $4\%$. We could then easily infer whether ionising photons at
$z\sim 2.5$ are typically produced by galaxies, quasars or both. This
does assume that the calibration of $\bFg$ is perfect, but given close
agreement between simulations and generic analytic arguments (Section
\ref{sec:params}), such an assumption is warranted.

There is currently disagreement over the precise photon mean free path
\citep[{\it e.g.},][]{Rudie13MFP,Prochaska13_MFP}, so one might allow
$\kappaHI$ to float. In this scenario, constraints on $b_j$ and
$\bar{n}$ degrade significantly (see $\times$ symbols in Figure
\ref{fig:fisher}) to $20\%$ errors on $b_j$ and $\simeq 60\%$ errors
on $\bar{n}$.
The right panel of Figure \ref{fig:fisher} shows degeneracies between
the parameters for this case, illustrating why advance knowledge is so
helpful: there is a near-complete three-way degeneracy between
$\kappaHI$, $\bar{n}$ and $b_j$. Increasing $\kappaHI$ brings the
effects of radiative transfer to smaller scales; this can be
counterbalanced by increasing the number density and reducing the
clustering of the sources. However a weak prior on $\kappaHI$ is
sufficient to break the degeneracy; for example, taking a $\kappaHI$
prior with standard deviation $\pm 20\%$, the forest gives estimates
accurate to $7\%$ ($b_j$) and $22\%$ ($\bar{n}$).

In the left panel of Figure~\ref{fig:fisher} we have also plotted (as
$+$ symbols) the forecasts for {\tt BOSS}. One obtains $\sim 15\%$
constraints on $b_j$ and $\bar{n}$; all the above caveats apply,
in that one needs a reasonable prior on $\kappaHI$ to achieve this.
The uppermost line (with $\circ$ symbols) shows {\tt BOSS} constraints
marginalized over an unknown broadband distortion, which the current
pipelines require; floating the distortion parameters
$a_{\ell,n}$ implies loss of traction on the astrophysics.
In other words the
astrophysical information is lost unless the broadband shape of $\xiF$
can be reconstructed accurately. Robust methods to calibrate the
quasar continuum are therefore required
\citep[{\it e.g.},][]{Paris11continuum,Lee12_continuum}.

We found that constraints on the BAO peak location are not degraded by
marginalising over the astrophysical parameters we consider (compared
against an ideal case where radiative distortions are fixed or
artificially switched off). For instance $\alpha$, the BAO position,
is constrained to $\pm 2.8\,\%$ with {\tt BOSS}; we verified this is
exactly the same constraint as obtained in the ideal, undistorted
case.  This applies to an isotropic rescaling, but different
cosmological information is available from the rescaling along and
perpendicular to the line of sight
\citep{BOSS_Slosar13,Delubac14_Lya_BOSSDR11}. When constraining these
directions separately, anisotropic distortions can mix the moments of
the correlation function so leading to broadband distortions. The main
effect is to slightly degrade the astrophysical constraints. Taking the
{\tt DESI} case, for example, we find that the relative uncertainty on
$b_j$ rises from $4\%$ with isotropic scaling to $6\%$ when the
scaling along the two directions is independent.

%

\section{Conclusions}\label{sec:conclusions}

We have investigated the potential of present and planned surveys of
the large-scale Lyman-$\alpha$ forest to reveal properties of the
objects producing ionising photons at $z\simeq 2.5$.  We find that,
provided pipelines can be developed where the broadband calibration
distortions are largely eliminated or fully characterized, there are
excellent prospects for deriving meaningful data on population number
density $\bar{n}$ and clustering strength $b_j$. These constraints can
in turn be used to discriminate between galaxy- and quasar-dominated
scenarios; or compared against a more sophisticated model for the
origin of photons in which $\bar{n}$ and $b_j$ become weighted
population averages. Such an approach would be highly complementary to
constructing luminosity functions for high-redshift galaxies and
quasars, since it traces all emission rather than just that coming
from the brightest objects -- and automatically folds in the effect of
varying escape fractions.

We have not yet investigated the additional power that would come from
cross-correlation studies
\cite[\textit{e.g.},][]{FontRibera_BOSS_QSO_Lya_X_2013}, nor have we
investigated how splitting the results into redshift bins could give
constraints on evolution.  The results we have presented assume that
the small-scale nonlinearities can be simulated well enough to
consider the radiative flux bias $\bFg$ known, and that $\sigma_8$ can
be derived from other datasets; without this calibration, one would
instead have to estimate ratios such as $b_j/\bu$. We have ignored
systematics such as residual metal-line contamination
\citep{SlosarLyaMetalContam14} and large-scale temperature
fluctuations \citep{McQuinn11_LyA_fluctuations,Gontcho14}. It is
currently unclear to what extent these will be degenerate with each
other and with the constraints we seek -- a campaign of simulations
and analytic work to understand how various effects and approximations
interact with each other is required. But the basic result that
astrophysics distorts the large scale forest fluctuations in useful,
measurable ways will survive, even if the degeneracies are somewhat
more complex than we can currently model.

\pagebreak

\section*{Acknowledgements}
We are grateful to the anonymous referee for many clarifications
and to George Becker, Jamie Bolton, Andreu Font-Ribera, Nick Kaiser,
Jordi Miralda-Escud{\'e}, Daniela Saadeh, Brian Siana and Risa
Wechsler for helpful discussions. AP is supported by a Royal Society
University Research Fellowship. SB is supported by the National
Science Foundation grant number AST-0907969, the W.M. Keck Foundation
and the Institute for Advanced Study.  HVP is supported by STFC and
the European Research Council under the European Community’s Seventh
Framework Programme (FP7/2007- 2013) / ERC grant agreement no
306478-CosmicDawn. LV is supported by European Research Council under
the European Communities Seventh Framework Programme grant
FP7-IDEAS-Phys.LSS and acknowledges Mineco grant FPA2011-29678-
C02-02. Some numerical results were derived with the {\sc
  Pynbody} framework \citep{2013ascl.soft05002P}.


\begin{thebibliography}{}
\expandafter\ifx\csname natexlab\endcsname\relax\def\natexlab#1{#1}\fi

\bibitem[{{Albrecht} {et~al.}(2006){Albrecht}, {Bernstein}, {Cahn}, {Freedman},
  {Hewitt}, {Hu}, {Huth}, {Kamionkowski}, {Kolb}, {Knox}, {Mather}, {Staggs},
  \& {Suntzeff}}]{DETF06}
{Albrecht}, A., {Bernstein}, G., {Cahn}, R., {et~al.} 2006, Report of the Dark
  Energy Task Force, astro-ph/0609591

\bibitem[{{Bajtlik} {et~al.}(1988){Bajtlik}, {Duncan}, \&
  {Ostriker}}]{Batjlik88_proximity}
{Bajtlik}, S., {Duncan}, R.~C., \& {Ostriker}, J.~P. 1988, \apj, 327, 570

\bibitem[{{Bird} {et~al.}(2011){Bird}, {Peiris}, {Viel}, \&
  {Verde}}]{Bird11_Lya}
{Bird}, S., {Peiris}, H.~V., {Viel}, M., \& {Verde}, L. 2011, \mnras, 413, 1717

\bibitem[{{Bower} {et~al.}(1993){Bower}, {Coles}, {Frenk}, \&
  {White}}]{Bower93_cooperativeGform}
{Bower}, R.~G., {Coles}, P., {Frenk}, C.~S., \& {White}, S.~D.~M. 1993, \apj,
  405, 403

\bibitem[{{Busca} {et~al.}(2013){Busca}, {Delubac}, {Rich}, {Bailey},
  {Font-Ribera}, {Kirkby}, {Le Goff}, {Pieri}, {Slosar}, {Aubourg}, {Bautista},
  {Bizyaev}, {Blomqvist}, {Bolton}, {Bovy}, {Brewington}, {Borde}, {Brinkmann},
  {Carithers}, {Croft}, {Dawson}, {Ebelke}, {Eisenstein}, {Hamilton}, {Ho},
  {Hogg}, {Honscheid}, {Lee}, {Lundgren}, {Malanushenko}, {Malanushenko},
  {Margala}, {Maraston}, {Mehta}, {Miralda-Escud{\'e}}, {Myers}, {Nichol},
  {Noterdaeme}, {Olmstead}, {Oravetz}, {Palanque-Delabrouille}, {Pan},
  {P{\^a}ris}, {Percival}, {Petitjean}, {Roe}, {Rollinde}, {Ross}, {Rossi},
  {Schlegel}, {Schneider}, {Shelden}, {Sheldon}, {Simmons}, {Snedden},
  {Tinker}, {Viel}, {Weaver}, {Weinberg}, {White}, {Y{\`e}che}, \&
  {York}}]{2013_Busca_BOSS_Lya_BAO}
{Busca}, N.~G., {Delubac}, T., {Rich}, J., {et~al.} 2013, \aap, 552, A96

\bibitem[{{Calverley} {et~al.}(2011){Calverley}, {Becker}, {Haehnelt}, \&
  {Bolton}}]{Calverley_11_proximity}
{Calverley}, A.~P., {Becker}, G.~D., {Haehnelt}, M.~G., \& {Bolton}, J.~S.
  2011, \mnras, 412, 2543

\bibitem[{{Croft}(2004)}]{Croft04_LyA_large_radiation}
{Croft}, R. A.~C. 2004, \apj, 610, 642

\bibitem[{{Croft} {et~al.}(1999){Croft}, {Weinberg}, {Pettini}, {Hernquist}, \&
  {Katz}}]{Croft99}
{Croft}, R.~A.~C., {Weinberg}, D.~H., {Pettini}, M., {Hernquist}, L., \&
  {Katz}, N. 1999, \apj, 520, 1

\bibitem[{{Dall'Aglio} {et~al.}(2008){Dall'Aglio}, {Wisotzki}, \&
  {Worseck}}]{Dallaglio_08_proximity}
{Dall'Aglio}, A., {Wisotzki}, L., \& {Worseck}, G. 2008, \aap, 491, 465

\bibitem[{{Dawson} {et~al.}(2013){Dawson}, {Schlegel}, {Ahn}, {Anderson},
  {Aubourg}, {Bailey}, {Barkhouser}, {Bautista}, {Beifiori}, {Berlind},
  {Bhardwaj}, {Bizyaev}, {Blake}, {Blanton}, {Blomqvist}, {Bolton}, {Borde},
  {Bovy}, {Brandt}, {Brewington}, {Brinkmann}, {Brown}, {Brownstein}, {Bundy},
  {Busca}, {Carithers}, {Carnero}, {Carr}, {Chen}, {Comparat}, {Connolly},
  {Cope}, {Croft}, {Cuesta}, {da Costa}, {Davenport}, {Delubac}, {de Putter},
  {Dhital}, {Ealet}, {Ebelke}, {Eisenstein}, {Escoffier}, {Fan}, {Filiz Ak},
  {Finley}, {Font-Ribera}, {G{\'e}nova-Santos}, {Gunn}, {Guo}, {Haggard},
  {Hall}, {Hamilton}, {Harris}, {Harris}, {Ho}, {Hogg}, {Holder}, {Honscheid},
  {Huehnerhoff}, {Jordan}, {Jordan}, {Kauffmann}, {Kazin}, {Kirkby}, {Klaene},
  {Kneib}, {Le Goff}, {Lee}, {Long}, {Loomis}, {Lundgren}, {Lupton}, {Maia},
  {Makler}, {Malanushenko}, {Malanushenko}, {Mandelbaum}, {Manera}, {Maraston},
  {Margala}, {Masters}, {McBride}, {McDonald}, {McGreer}, {McMahon}, {Mena},
  {Miralda-Escud{\'e}}, {Montero-Dorta}, {Montesano}, {Muna}, {Myers},
  {Naugle}, {Nichol}, {Noterdaeme}, {Nuza}, {Olmstead}, {Oravetz}, {Oravetz},
  {Owen}, {Padmanabhan}, {Palanque-Delabrouille}, {Pan}, {Parejko},
  {P{\^a}ris}, {Percival}, {P{\'e}rez-Fournon}, {P{\'e}rez-R{\`a}fols},
  {Petitjean}, {Pfaffenberger}, {Pforr}, {Pieri}, {Prada}, {Price-Whelan},
  {Raddick}, {Rebolo}, {Rich}, {Richards}, {Rockosi}, {Roe}, {Ross}, {Ross},
  {Rossi}, {Rubi{\~n}o-Martin}, {Samushia}, {S{\'a}nchez}, {Sayres}, {Schmidt},
  {Schneider}, {Sc{\'o}ccola}, {Seo}, {Shelden}, {Sheldon}, {Shen}, {Shu},
  {Slosar}, {Smee}, {Snedden}, {Stauffer}, {Steele}, {Strauss}, {Streblyanska},
  {Suzuki}, {Swanson}, {Tal}, {Tanaka}, {Thomas}, {Tinker}, {Tojeiro},
  {Tremonti}, {Vargas Maga{\~n}a}, {Verde}, {Viel}, {Wake}, {Watson}, {Weaver},
  {Weinberg}, {Weiner}, {West}, {White}, {Wood-Vasey}, {Yeche}, {Zehavi},
  {Zhao}, \& {Zheng}}]{2013Dawson_BOSS_overview}
{Dawson}, K.~S., {Schlegel}, D.~J., {Ahn}, C.~P., {et~al.} 2013, \aj, 145, 10

\bibitem[{{Delubac} {et~al.}(2014){Delubac}, {Bautista}, {Busca}, {Rich},
  {Kirkby}, {Bailey}, {Font-Ribera}, {Slosar}, {Lee}, {Pieri}, {Hamilton},
  {Aubourg}, {Blomqvist}, {Bovy}, {Brinkmann}, {Carithers}, {Dawson},
  {Eisenstein}, {Kneib}, {Le Goff}, {Margala}, {Miralda-Escud{\'e}}, {Myers},
  {Nichol}, {Noterdaeme}, {O'Connell}, {Olmstead}, {Palanque-Delabrouille},
  {P{\^a}ris}, {Petitjean}, {Ross}, {Rossi}, {Schlegel}, {Schneider},
  {Weinberg}, {Y{\`e}che}, \& {York}}]{Delubac14_Lya_BOSSDR11}
{Delubac}, T., {Bautista}, J.~E., {Busca}, N.~G., {et~al.} 2014, A\&A submitted, arXiv:1404.1801

\bibitem[{{Font-Ribera} {et~al.}(2013){Font-Ribera}, {Arnau},
  {Miralda-Escud{\'e}}, {Rollinde}, {Brinkmann}, {Brownstein}, {Lee}, {Myers},
  {Palanque-Delabrouille}, {P{\^a}ris}, {Petitjean}, {Rich}, {Ross},
  {Schneider}, \& {White}}]{FontRibera_BOSS_QSO_Lya_X_2013}
{Font-Ribera}, A., {Arnau}, E., {Miralda-Escud{\'e}}, J., {et~al.} 2013, \jcap,
  5, 18

\bibitem[{{Gontcho} {et~al.}(2014){Gontcho}, {Miralda-Escud{\'e}}, \&
  {Busca}}]{Gontcho14}
{Gontcho}, S.~G.~A., {Miralda-Escud{\'e}}, J., \& {Busca}, N.~G. 2014, MNRAS, accepted:1404.7425

\bibitem[{{Gunn} \& {Peterson}(1965)}]{1965ApJ...142.1633G}
{Gunn}, J.~E., \& {Peterson}, B.~A. 1965, \apj, 142, 1633

\bibitem[{{Hamilton}(1998)}]{Hamilton98_RSDreview}
{Hamilton}, A.~J.~S. 1998, in Astrophysics and Space Science Library, Vol. 231,
  The Evolving Universe, ed. D.~{Hamilton}, 185

\bibitem[{{Ir{\v s}i{\v c}} \& {Slosar}(2014)}]{SlosarLyaMetalContam14}
{Ir{\v s}i{\v c}}, V., \& {Slosar}, A. 2014, \prd, 89, 107301

\bibitem[{{Kaiser}(1987)}]{1987MNRAS.227....1K}
{Kaiser}, N. 1987, \mnras, 227, 1

\bibitem[{{Kirkby} {et~al.}(2013){Kirkby}, {Margala}, {Slosar}, {Bailey},
  {Busca}, {Delubac}, {Rich}, {Bautista}, {Blomqvist}, {Brownstein},
  {Carithers}, {Croft}, {Dawson}, {Font-Ribera}, {Miralda-Escud{\'e}}, {Myers},
  {Nichol}, {Palanque-Delabrouille}, {P{\^a}ris}, {Petitjean}, {Rossi},
  {Schlegel}, {Schneider}, {Viel}, {Weinberg}, \& {Y{\`e}che}}]{Kirkby13Lya}
{Kirkby}, D., {Margala}, D., {Slosar}, A., {et~al.} 2013, \jcap, 3, 24

\bibitem[{{Kollmeier} {et~al.}(2003){Kollmeier}, {Weinberg}, {Dav{\'e}}, \&
  {Katz}}]{Kollmeier03_LyA_LBG_connection}
{Kollmeier}, J.~A., {Weinberg}, D.~H., {Dav{\'e}}, R., \& {Katz}, N. 2003,
  \apj, 594, 75

\bibitem[{{Lee}(2012)}]{Lee12_continuum}
{Lee}, K.-G. 2012, \apj, 753, 136

\bibitem[{{Levi} {et~al.}(2013){Levi}, {Bebek}, {Beers}, {Blum}, {Cahn},
  {Eisenstein}, {Flaugher}, {Honscheid}, {Kron}, {Lahav}, {McDonald}, {Roe},
  {Schlegel}, \& {representing the DESI collaboration}}]{Levi13_DESI}
{Levi}, M., {Bebek}, C., {Beers}, T., {et~al.} 2013, Snowmass white
paper,
  arXiv:1308.0847

\bibitem[{Lewis {et~al.}(2000)Lewis, Challinor, \& Lasenby}]{Lewis:1999bs}
Lewis, A., Challinor, A., \& Lasenby, A. 2000, Astrophys. J., 538, 473

\bibitem[{{McDonald}(2003)}]{McDonald03}
{McDonald}, P. 2003, \apj, 585, 34

\bibitem[{{McDonald} \&
  {Eisenstein}(2007)}]{McDonaldEisenstein07_LyABAOforecast}
{McDonald}, P., \& {Eisenstein}, D.~J. 2007, \prd, 76, 063009

\bibitem[{{McDonald} {et~al.}(2000){McDonald}, {Miralda-Escud{\'e}}, {Rauch},
  {Sargent}, {Barlow}, {Cen}, \& {Ostriker}}]{McDonald00}
{McDonald}, P., {Miralda-Escud{\'e}}, J., {Rauch}, M., {et~al.} 2000, \apj,
  543, 1

\bibitem[{{McDonald} {et~al.}(2005){McDonald}, {Seljak}, {Cen}, {Bode}, \&
  {Ostriker}}]{McDonald05}
{McDonald}, P., {Seljak}, U., {Cen}, R., {Bode}, P., \& {Ostriker}, J.~P. 2005,
  \mnras, 360, 1471

\bibitem[{{McDonald} {et~al.}(2006){McDonald}, {Seljak}, {Burles}, {Schlegel},
  {Weinberg}, {Cen}, {Shih}, {Schaye}, {Schneider}, {Bahcall}, {Briggs},
  {Brinkmann}, {Brunner}, {Fukugita}, {Gunn}, {Ivezi{\'c}}, {Kent}, {Lupton},
  \& {Vanden Berk}}]{McDonald_SDSS_1DLyA_Pk_2006}
{McDonald}, P., {Seljak}, U., {Burles}, S., {et~al.} 2006, \apjs, 163, 80

\bibitem[{{McQuinn} {et~al.}(2011{\natexlab{a}}){McQuinn}, {Hernquist}, {Lidz},
  \& {Zaldarriaga}}]{McQuinn11_LyA_fluctuations}
{McQuinn}, M., {Hernquist}, L., {Lidz}, A., \& {Zaldarriaga}, M.
  2011{\natexlab{a}}, \mnras, 415, 977

\bibitem[{{McQuinn} {et~al.}(2011{\natexlab{b}}){McQuinn}, {Oh}, \&
  {Faucher-Gigu{\`e}re}}]{McQuinn_11_LyLimit}
{McQuinn}, M., {Oh}, S.~P., \& {Faucher-Gigu{\`e}re}, C.-A. 2011{\natexlab{b}},
  \apj, 743, 82

\bibitem[{{McQuinn} \& {White}(2011)}]{McQuinnWhite11_LyA_estimators}
{McQuinn}, M., \& {White}, M. 2011, \mnras, 415, 2257

\bibitem[{{Meiksin} \& {White}(2004)}]{Meiskin04_LyA_radiation}
{Meiksin}, A., \& {White}, M. 2004, \mnras, 350, 1107

\bibitem[Murdoch et al.(1986)]{1986ApJ...309...19M} Murdoch, H.~S.,
Hunstead, R.~W., Pettini, M., \& Blades, J.~C.\ 1986, \apj, 309, 19

\bibitem[{{P{\^a}ris} {et~al.}(2011){P{\^a}ris}, {Petitjean}, {Rollinde},
  {Aubourg}, {Busca}, {Charlassier}, {Delubac}, {Hamilton}, {Le Goff},
  {Palanque-Delabrouille}, {Peirani}, {Pichon}, {Rich}, {Vargas-Maga{\~n}a}, \&
  {Y{\`e}che}}]{Paris11continuum}
{P{\^a}ris}, I., {Petitjean}, P., {Rollinde}, E., {et~al.} 2011, \aap, 530, A50

\bibitem[{{Planck Collaboration}(2013)}]{Planck13_CosmoPar}
{Planck Collaboration}. 2013, A\&A accepted, arXiv:1303.5076

\bibitem[{{Pontzen}(2014)}]{Pontzen14_BAObias}
{Pontzen}, A. 2014, \prd, 89, 083010

\bibitem[{{Pontzen} {et~al.}(2013){Pontzen}, {Roskar}, {Stinson}, \&
  {Woods}}]{2013ascl.soft05002P}
{Pontzen}, A., {Roskar}, R., {Stinson}, G., \& {Woods}, R. 2013, {pynbody:
  N-Body/SPH analysis for python}, astrophysics Source Code Library
  ascl:1305.002, ascl:1305.002

\bibitem[{{Prochaska} {et~al.}(2014){Prochaska}, {Madau}, {O'Meara}, \&
  {Fumagalli}}]{Prochaska13_MFP}
{Prochaska}, J.~X., {Madau}, P., {O'Meara}, J.~M., \& {Fumagalli}, M. 2014,
  \mnras, 438, 476

\bibitem[{{Rudie} {et~al.}(2013){Rudie}, {Steidel}, {Shapley}, \&
  {Pettini}}]{Rudie13MFP}
{Rudie}, G.~C., {Steidel}, C.~C., {Shapley}, A.~E., \& {Pettini}, M. 2013,
  \apj, 769, 146

\bibitem[{{Seljak}(2012)}]{Seljak12LyaBias}
{Seljak}, U. 2012, \jcap, 3, 4

\bibitem[{{Slosar} {et~al.}(2011){Slosar}, {Font-Ribera}, {Pieri}, {Rich}, {Le
  Goff}, {Aubourg}, {Brinkmann}, {Busca}, {Carithers}, {Charlassier},
  {Cort{\^e}s}, {Croft}, {Dawson}, {Eisenstein}, {Hamilton}, {Ho}, {Lee},
  {Lupton}, {McDonald}, {Medolin}, {Muna}, {Miralda-Escud{\'e}}, {Myers},
  {Nichol}, {Palanque-Delabrouille}, {P{\^a}ris}, {Petitjean}, {Pi{\v s}kur},
  {Rollinde}, {Ross}, {Schlegel}, {Schneider}, {Sheldon}, {Weaver}, {Weinberg},
  {Yeche}, \& {York}}]{BOSS_Slosar11}
{Slosar}, A., {Font-Ribera}, A., {Pieri}, M.~M., {et~al.} 2011, \jcap, 9, 1

\bibitem[{{Slosar} {et~al.}(2013){Slosar}, {Ir{\v s}i{\v c}}, {Kirkby},
  {Bailey}, {Busca}, {Delubac}, {Rich}, {Aubourg}, {Bautista}, {Bhardwaj},
  {Blomqvist}, {Bolton}, {Bovy}, {Brownstein}, {Carithers}, {Croft}, {Dawson},
  {Font-Ribera}, {Le Goff}, {Ho}, {Honscheid}, {Lee}, {Margala}, {McDonald},
  {Medolin}, {Miralda-Escud{\'e}}, {Myers}, {Nichol}, {Noterdaeme},
  {Palanque-Delabrouille}, {P{\^a}ris}, {Petitjean}, {Pieri}, {Pi{\v s}kur},
  {Roe}, {Ross}, {Rossi}, {Schlegel}, {Schneider}, {Suzuki}, {Sheldon},
  {Seljak}, {Viel}, {Weinberg}, \& {Y{\`e}che}}]{BOSS_Slosar13}
{Slosar}, A., {Ir{\v s}i{\v c}}, V., {Kirkby}, D., {et~al.} 2013, \jcap, 4, 26

\end{thebibliography}

\end{document}